\documentclass[5p,times,procedia]{elsarticle}
\usepackage{graphicx}
\usepackage{svg}
\usepackage{comment}
\usepackage{paralist}
\usepackage{amsmath}
\usepackage[table]{xcolor}
\usepackage{tikz}
\definecolor{green}{RGB}{76, 187, 23} 
\definecolor{yellow}{RGB}{255,165,0}     
\definecolor{red}{RGB}{255,0,0}

\usepackage{tikz}
\usetikzlibrary{shapes}
\usetikzlibrary{arrows.meta}
\usepackage{subcaption}

\usepackage[resetlabels,labeled]{multibib}

\usepackage{todonotes}

\usepackage{ecrc}

\volume{00}

\firstpage{1}

\journalname{Procedia CIRP}

\runauth{C. Ellwein et al.}

\jid{trpro}

\usepackage{amssymb}

\biboptions{numbers}

\usepackage[figuresright]{rotating}

\usepackage[bookmarks=false]{hyperref}
    \hypersetup{colorlinks,
      linkcolor=blue,
      citecolor=blue,
      urlcolor=blue}

\begin{document}
\begin{frontmatter}



\dochead{CIRP Global Web Conference 2026 (CIRPe 2026)}%

\title{A Set-Theoretic Evaluation Framework for Assessing Asset Administration Shell Instances: Towards Comparability and Suitability}


\author[a]{Carsten Ellwein\corref{cor1}} 
\author[a]{David Dietrich} 
\author[b]{Rozana Cvitkovic}
\author[b]{\\Bastian Lang}
\author[a]{Jessica Roth}
\author[b]{Hansjoerg Tutsch}
\author[a]{Andreas Wortmann} 

\address[a]{Institute for Control Engineering of Machine Tools and Manufacturing Units (ISW), University of Stuttgart, Seidenstraße 36, 70174 Stuttgart, Germany}
\address[b]{Blue Yonder GmbH,Industriestraße 6, 70565 Stuttgart, Germany}

\cortext[cor1]{* Corresponding author. Tel.: +49 711 685-82424 {\it E-mail address:} carsten.ellwein@isw.uni-stuttgart.de}

\begin{abstract}
Asset Administration Shells (AAS) provide a standardized means of representing assets and their information in manufacturing and increasingly serve as a basis for software services. However, different AAS instances vary in structure, content, and degree of completion, making it difficult to determine whether a given AAS is suitable for a specific application. This paper presents two complementary methods to support the comparison and application-oriented assessment of AAS. First, set-theoretic operations are employed to compare AAS models, enabling the identification of common, missing, and differing submodels and parameters. Second, an AAS suitability model assesses the conformity of an AAS to the requirements of a specific use case. The assessment considers structural conformity, semantic consistency, cardinality, and specification conformity and can be performed either against a reference AAS or a set of required SemanticIDs. A suitability value is derived from the identified deviations and is complemented by a detailed report of missing or non-conforming information. The proposed approach support practitioners and researchers in the comparison of evolving AAS and provide application-specific information on their suitability for manufacturing software services.
\end{abstract}


\end{frontmatter}

\section{Introduction}
\label{sec:Introduction}
Digital twins are increasingly becoming the technological backbone to improve the understanding, design, operation, and management of cyber-physical systems (CPS)~\cite{DJR+22}.
Although the software engineering community is primarily responsible for developing the key technologies that underpin DTs~\cite{FHM+23}, manufacturing continues to be one of the primary application areas in which their capabilities are explored~\cite{chen2025}.
Within manufacturing, a digital twin is typically defined as a virtual representation of a cyber-physical production system (CPPS) \cite{EBC+21}, instantiated and executed within a software environment~\cite{munoz2023conceptual}.
Individual digital twins are tailored to specific objectives with respect to their corresponding physical systems~\cite{EBC+21}, e.g., analysis~\cite{827}, control~\cite{1152}, or behavior forecast~\cite{splettstosser2023self}. 

The Asset Administration Shell (AAS) is becoming a central technology for representing and implementing digital twins~\cite{Neubauer_2023} in manufacturing and has already been employed to describe products~\cite{chen2025}, processes~\cite{dietrich2024}, and resources~\cite{frick2024software}.
The AAS is designed to serve as the authoritative and instantiated digital representation of any asset throughout its life cycle~\cite{bader2022details}.
Consequently, the AAS is therefore utilized and further continuously advanced by various stakeholders~--~partly simultaneously, partly in a collaborative manner.
This, along with the use of multiple submodels, increases the risk of ambiguity within the AAS. 
Set theory has previously been used to formally compare data models, e.g. by defining relative information capacity between database schemata via set-based mappings between their instances \cite{hull1986schemata}, including more recent set-theoretic treatments of the AAS itself \cite{bader2019semantic}. 
However, existing approaches to AAS comparability rely on informal or purely syntactic matching rather than a formal set-theoretic basis capable of guaranteeing precise notions such as equivalence and subsumption.
The first part of the paper therefore establishes the methodological foundations necessary for a systematic and coherent analysis of AAS instances.

The AAS is increasingly employed not merely as a representational construct but as a foundational framework for the provision of software-based services~\cite{ellwein2025container, ellwein2026software}.
This raises the challenge of comparing different AAS to identify the appropriate data source for the software component to be implemented and to ensure that the required information is available in the expected format. 
An option is to develop and apply a dedicated maturity model~\cite{Ellwein2026nier}. 
However, maturity does not necessarily indicate suitability for a specific use case. 
The approach discussed in the second part of the paper therefore is a reference-based suitability assessment, in which the application developer provides a reference or template AAS that specifies the minimum requirement profile.

In the following, \autoref{sec:RelatedWork} presents the state of the art on AAS. Subsequently, \autoref{sec:Comparability} introduces a concept of mutual comparability for AAS models grounded in set theory, before \autoref{sec:Capability} develops the AAS suitability model. 
Next, \autoref{sec:Exemplification} demonstrates how the proposed methods can be applied by means of an example. 
Building on these results, \autoref{sec:limitations} highlights limitations, before \autoref{sec:Conclusion} offers concluding remarks.

\section{Background}
\label{sec:RelatedWork}

Within the framework of Industry 4.0, any entity owned by an organization that adds value to the execution of processes is termed an asset~\cite{cavalieri2020asset}. 
Assets can be physical, such as production machinery, workpieces, or even the entire manufacturing plant, but they can also be non-physical, including models used to represent machine behavior, software, or licenses~\cite{Frysak.2018}.

The AAS~\cite{wei2019review} is increasingly being recognized as a potential standard for modeling and implementing digital twins~\cite{Neubauer_2023}, and has been proposed as \textit{the definition and representation} layer in ISO 23247.
Its evolution is being advanced by the International Digital Twin Association (IDTA)~\cite{bader2019semantic}, an industry-focused organization affiliated with the Association of German Mechanical Engineering Companies (VDMA). 

Because the AAS holds relevant information covering the entire life cycle of an asset, it must be capable of depicting a wide range of content, including properties, modeled functions, parameters, summaries of integrated components, data generated during manufacturing or simulation, and descriptive details such as usage guidelines and technical specifications. 
Consequently, it needs the ability to store or reference heterogeneous data types~\cite{cavalieri2020asset}.

The structure of the AAS is defined by a metamodel (cf.~\cite{bader2022details}). 
At the top level, the \textit{AssetAdministrationShell} class represents the complete AAS of a given asset.
It may contain multiple submodels, represented by the \textit{Submodel} class, each capturing particular characteristics of the asset. 
The individual attributes that constitute a submodel — such as properties and files — are represented by the abstract class \textit{SubmodelElement}.
From this, the abstract class \textit{DataElement} is derived and serves as a base class for further concrete specializations. The classes \textit{Property}, \textit{Range}, \textit{File}, and \textit{ReferenceElement} describe different categories of asset attributes. 
\textit{Property} encapsulates a pair of information consisting of a single value and its data type, whereas \textit{Range} defines two values that share a common data type. 
\textit{File} specifies the type and location of a file, while \textit{ReferenceElement} establishes a logical reference either to another element within the same AAS or to an element in a different AAS.

To ensure consistency and interoperability, the IDTA provides so-called submodel templates. These templates are standardized as well and are publicly available through the Content Hub\footnote{IDTA Content Hub: \url{https://industrialdigitaltwin.org/en/content-hub/submodels}}.
Each parameter in a submodel template is specified by an identifier, its semantic definition, and a representative example.  
The semantic definition is harmonized with established dictionaries, such as the ECLASS\footnote{ECLASS: \url{https://eclass.eu/en/eclass-standard}} reference data standard for the unequivocal description of products and services, and the IEC Common Data Dictionary, and is referred to as \textit{SemanticID}.  

\section{AAS Comparability}
\label{sec:Comparability}
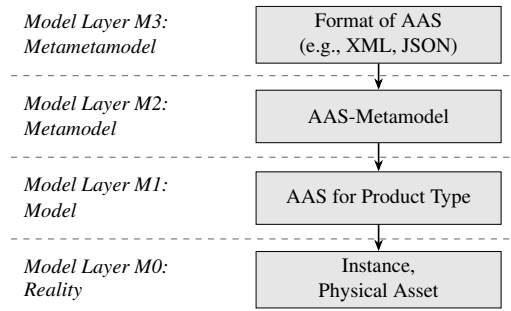
\begin{figure}
\centering
    \resizebox{0.75\columnwidth}{!}{%
\begin{tikzpicture}[
    layerline/.style={gray, dashed, line width=0.6pt},
    box/.style={
        draw=black,
        fill=gray!20,
        minimum width=4.2cm,
        minimum height=0.9cm,
        align=center
    },
    arrow/.style={
        -{Stealth[length=2mm]},
        line width=0.8pt,
        black
    }
]

\node[anchor=west] at (0,4.8) {\textit{Model Layer M3:}};
\node[anchor=west] at (0,4.4) {\textit{Metametamodel}};

\node[anchor=west] at (0,3.4) {\textit{Model Layer M2:}};
\node[anchor=west] at (0,3.0) {\textit{Metamodel}};

\node[anchor=west] at (0,2.0) {\textit{Model Layer M1:}};
\node[anchor=west] at (0,1.6) {\textit{Model}};

\node[anchor=west] at (0,0.6) {\textit{Model Layer M0:}};
\node[anchor=west] at (0,0.2) {\textit{Reality}};

\node[box] (m3) at (6.2,4.6) {Format of AAS\\(e.g., XML, JSON)};
\node[box] (m2) at (6.2,3.2) {AAS-Metamodel};
\node[box] (m1) at (6.2,1.8) {AAS for Product Type};
\node[box] (m0) at (6.2,0.4) {Instance,\\Physical Asset};

\draw[arrow] (m3) -- (m2);
\draw[arrow] (m2) -- (m1);
\draw[arrow] (m1) -- (m0);

\draw[layerline] (-0.1,3.9) -- (8.5,3.9);
\draw[layerline] (-0.1,2.5) -- (8.5,2.5);
\draw[layerline] (-0.1,1.1) -- (8.5,1.1);

\end{tikzpicture}%
}
\caption{The four layers of an AAS, based on the hierarchical metamodel layout \cite{flatscher2002}}
\label{fig:FourLayer}
\end{figure}

This chapter describes the comparability of AAS instances. 
The focus of this analysis is the AAS model.
Since AASs are complex constructs, it is necessary to define the level at which the comparison is to be made. 
The four-layer hierarchical metamodel layout~\cite{flatscher2002}, which is also used by the Object Management Group (OMG) and the Electronic Industries Alliance / Computer-Assisted software engineering (CASE) Data Interchange Format (EIA/CDIF), is applied to the AAS. 
In this hierarchical layout, the upper level specifies the lower level. 
This means that the lower level can be regarded as an instance of the upper level. 
The four levels are formally defined as M3 - Metametamodel, M2 - Metamodel, M1 - User Model, and M0 - User Object \cite{alvarez2001}.
Thus, M0 describes the current system and the objects that currently exist. 
M1 represents these objects as models or classes with defined attributes. 
M1 is the classification of M0 and, conversely, the objects are instances of M1. 
M2 describes the modeling language, i.e., the process used to create models. 
M2 thus categorizes the models from M1. 
M3, as a metametamodel, describes the concepts used to create modeling languages in M2 \cite{fuentes2004}.
The layers applied to AAS can be seen in \autoref{fig:FourLayer}.

M3 describes the top level and thus the meta-metamodel of an AAS.
This includes general visualization and representation, as well as structuring languages such as JSON.
M2 contains the metamodel of an AAS, which includes general specifications and minimum requirements for an AAS.
M1 shows the AAS model for a product type, such as an AAS for a milling machine with the corresponding submodels.
M0 describes a specific instance of a physical asset, that is, real values for the parameters within the AAS submodels.
The question now arises as to which level AASs should be compared. 
M3 and M2 are standardized for AASs in order to precisely define which construct is already an AAS and which is not. 
This causes these levels to be identical when two AASs are compared to each other. 
Therefore, a comparison is only truly informative at the M1 and M0 levels.

The purpose of comparing two AASs is to identify similarities and differences between two physical assets. 
To do this, individual values within models would have to be compared. 
The same model must be present in both AASs in order to be able to compare individual data points within the models in the subsequent step. 
As a result, the comparison of AASs must first be carried out at the M1 level in order to do a meaningful comparison at the M0 level.
In order to compare AASs, the principles of set theory are used. 

The AAS constitutes a set, the submodels representing elements of this set. 
For demonstration purposes, three AASs containing several submodels are defined and compared in pairs.
\begin{equation}
A = \{1,2\} \text{, }
B = \{1,2,3\} \text{, }
C = \{1,3,4\}
\end{equation}
These AASs consist of submodels that can be identical or different.
For illustrative purposes, the submodels are represented numerically.
The numbers thus represent an identification number for a submodel.
The cardinality of an AAS is described by the number of its submodels.
In the example used here, this corresponds to:
\begin{equation}
\|A\| = 2  \text{, }
\|B\| = \|C\| = 3
\end{equation}
An AAS without submodels can be noted by an empty set.
\begin{equation}
D = \{\} = \emptyset
\end{equation}
An AAS consists of submodels, which in turn consist of empty and filled parameters with real values.
Submodels within an AAS are only valid if all mandatory parameters are filled in. 
The optional parameters may differ in terms of the percentage degree of fulfillment, that is, how many optional parameters are filled in within a submodel. 
Differences in the selection of the respective optional parameters are also possible. 


An AAS can be a subset of another AAS. 
This can occur, for example, when an AAS acquires additional submodels over time compared to its previous AAS version. 
Another use case would be the inheritance of models. 
If an AAS as the parent generation passes on some of its submodels to the child generation of AASs, or if the parent AAS transfers all of its submodels and the children also hold additional submodels, then this can be represented with subsets.
In the example introduced, A would be a subset of B but not of C. 
\begin{equation}
A \subseteq B  \text{, }
A \not\subseteq C
\end{equation}
Backward compatibility corresponds to $A \subseteq B$, and its absence to a non-trivial intersection with both difference sets non-empty.

The intersection between two AASs describes the common submodels. 
This makes it easy to determine the similarities between the AASs at the M1 model level.
If two AASs do not have any models in common, then they are disjunctive with respect to each other.
The intersection is then described with an empty set.
In the example described here, the intersections would look as follows.
\begin{equation}
A \cap B = \{ 1,2 \}  \text{, }
A \cap C = \{ 1 \}  \text{, }
C \cap B = \{ 1,3 \} 
\end{equation}

The union set describes an OR relationship between two AASs. 
The model must appear in either A or B or in both to be part of the set. 
This construct can be useful for obtaining an overview of which models are in use system-wide. 
This set can be used, for example, during system setup to obtain an overview of which models need to be implemented.
\begin{equation}
A \cup B = \{ 1,2,3 \}  \text{, }
A \cup C = \{ 1,2,3,4 \}  \text{, }
C \cup B = \{ 1,2,3,4 \} 
\end{equation}

Another way to compare AASs is using the difference set. 
For two AASs, A and B, the difference set describes all submodels that occur exclusively in A but not in B. 
This makes it possible, for example, to identify the differences of evolving AASs compared to the previous version. 
A typical use case would be the description of the version differences for the release documentation if submodels were added or removed. 
Applied to the example set, the difference sets are as follows.
The comparison of AASs B and C is skipped here.
\begin{equation}
A \backslash B = \emptyset  \text{, }
B \backslash A = \{ 3 \}  \text{, }
A \backslash C = \{ 2\}  \text{, }
C \backslash A = \{ 3,4 \} 
\end{equation}

Once a syntactic identity has been determined, i.e., a submodel exists in both AASs, a semantic comparison can be performed at the M0 level. 
In principle, the same set theory methods used for M1 can be applied to M0 as well.
To obtain a semantic statement, the real values assigned to the parameters of a submodel must be compared. 
With the help of the set theory, differences and similarities between the values can be determined and conclusions can be drawn about the two AASs being compared.
The submodel represents a set, and the values of the parameters represent the elements.
A submodel is termed empty if no parameters are filled with values.


Set-theoretic relations provide a formal basis for structured comparison. A submodel can be characterized as a subset of another submodel when the latter introduces additional parameters, while modifications of parameter values lead to non-trivial intersections rather than strict subset relations.  
In the context of evolving metamodels, intersections may also arise when backward compatibility is not preserved. In this case, intersections represent the set of shared parameters for which the corresponding values are identical, whereas disjoint submodels have no parameters with matching values in common.

The union operation aggregates all parameters and their associated values from both submodels, while the symmetric difference isolates those parameters that are either unique to one submodel or shared but assigned differing values, thus enabling explicit identification of discrepancies. Difference sets emphasize the features that are unique to a given submodel when compared to one or more reference submodels.

Complement operations support two distinct modes of comparative analysis. Structurally, they expose uninstantiated or missing parameters by referencing a designated complete parameter set. Semantically, they identify parameter values that are absent from, or deviate from, those in a broader contextual reference (e.g., a system-wide or fleet-wide baseline).

In summary, set theory can simplify the comparison of AASs by identifying both the differences and similarities between them using the various sets. Comparisons are carried out within levels M1 and M0 to examine syntactic and semantic identity.

\section{AAS Suitability Model}
\label{sec:Capability}
In addition to structural and organizational maturity aspects, the concrete applicability of an AAS is evaluated in specific application scenarios. Although conformity checks against specifications ensure syntactic and semantic correctness, they do not answer the central question from an application perspective: \textit{Is a specific AAS immediately suitable for use in a particular application?}
This chapter describes a methodical approach to testing the suitability of AAS. 
The aim is to produce a formalized assessment that provides immediate information on whether an AAS: (i) is fully usable, (ii) is usable with restrictions, or (iii) is not suitable due to missing information.

The suitability of an AAS for use is defined as: \textit{The degree of structural, semantic, and specification-compliant conformity between a given AAS and the requirements of a specific application.}
Formally, the test can be described as a comparison between two models: The AAS and a requirement model, given as list of defined semantic content or reference AAS -- the copy of a submodel template in which no values are specified, but entries are marked as required. 

The suitability check is carried out along four central dimensions that address different aspects of the conformity between a given AAS and the requirements of a specific application.

The first dimension concerns \textit{structural conformity}. If a reference administration shell is available, a structural model matching between the administration shell to be tested and this reference is performed. 
The comparison is performed at both the submodel level and the submodel elements contained therein. Methodologically, the comparison is similar to a structural diff of two JSON documents: It is analyzed whether all required submodels are present, whether the required elements exist, whether they are correctly embedded in the intended hierarchical structure, and whether the defined cardinality is adhered to. Missing mandatory submodels are considered critical deviations, as they usually prevent immediate usability of the administration shell for a specific use case. 
In contrast, missing optional elements are given less weight, as they restrict the application but do not necessarily exclude it.

The second dimension is \textit{semantic consistency}. As an alternative or supplement to structural comparison, suitability testing can be performed using SemanticIDs. In this case, the system checks whether there is a corresponding element in the AAS for each required SemanticID, whether the assignment is unique, and whether the associated data structure is consistent. The identity of SemanticID is considered a sufficient criterion for content consistency. There is no further examination of semantic hierarchies or ontological relationships, as the focus is on the availability of the specifically required information.

In addition to structural and semantic analysis, \textit{cardinality} and \textit{specification conformity} are also checked as third and fourth dimensions. For each relevant submodel and the elements it contains, the declared cardinality – for example, in the sense of 1..1 or 0..n – is compared with the actual number of instances and checked for conformity with the normative specifications of the IDTA. 
This involves checking whether mandatory elements are present, data types are used correctly, and — especially in the case of physical quantities — the specified units are in accordance with the specifications. 
This validation is implemented by a project-specific validator that converts the IDTA specifications into formalized test rules, enabling automated evaluation.
In the following, two comparison strategies are introduced.


The reference-based suitability assessment quantifies the conformance of an AAS to a requirement profile derived from the specific needs of the target use case.
This requirement profile is defined by the application developer in the form of a reference AAS, which specifies the minimum set of submodels and submodelelements required for the intended application.
Formally, we express this conformance using set difference operations.
Let $M_{req}$ denote the set of required elements specified by the reference AAS (e.g., required submodels, submodel elements, and their expected cardinality, type, and unit), with $\|M_{req}\|$ its cardinality. 
Suitability is then determined by:
\begin{equation}
S = 1 - \frac{\|D_{crit}\| + w \|D_{ncrit}\|}{\|M_{req}\|}
\label{eq:suit:ref}
\end{equation}
where $D_{crit} \subseteq M_{req}$ denotes the subset of critical deviations, i.e., required elements entirely missing from the evaluated AAS (e.g., a missing required submodel or mandatory element), and $D_{ncrit} \subseteq M_{req}$ denotes the subset of non-critical deviations, i.e., required elements that are present but do not fully conform to the reference (e.g., cardinality or unit discrepancies), with $D_{crit} \cap D_{ncrit} = \emptyset$.
The weighting factor $w \in [0,1]$ scales the non-critical deviations relative to critical ones.
Critical and non-critical deviations are specified in the reference AAS, whilst the weighting factor is defined by the user.
This method is particularly suitable for applications that are highly dependent on structure, such as production schedulers.

In the requirement-based suitability assessment, a list of required elements is defined using SemanticIDs.
This strategy is more robust against structural variations.
The suitability
\begin{equation}
S = \frac{\|E_{found}\|}{\|E_{req}\|}
\end{equation}
is thus calculated using found elements $E_{found}$ and required elements $E_{req}$.

The result of the check provides various information, such as: (i)~List of missing submodels, (ii)~list of missing or incorrect elements, (iii)~cardinality violations, (iv)~type and unit discrepancies, and (v)~general suitability value. 
This provides concrete transformation instructions to achieve the target suitability.
The suitability value $S$ is to be interpreted by the user in the context of their specific use case, as requirements on completeness and correctness vary depending on the intended application of the AAS.
Depending on the application of the AAS, requirements on completeness and correctness can vary, so the same suitability value may be regarded as sufficient in one scenario and insufficient in another.
The detailed breakdown of missing submodels, elements, cardinality violations, and discrepancies allows users to assess which deviations underlie the reported value and prioritize the transformation steps for their use case.
Critical deviations, such as missing required submodels, are additionally reported explicitly in the detailed breakdown, ensuring they remain visible to the user even when the aggregated suitability value alone might suggest high usability.



\section{Exemplification}
\label{sec:Exemplification}

\begin{figure}
    \centering
\includegraphics[width=0.7\columnwidth]{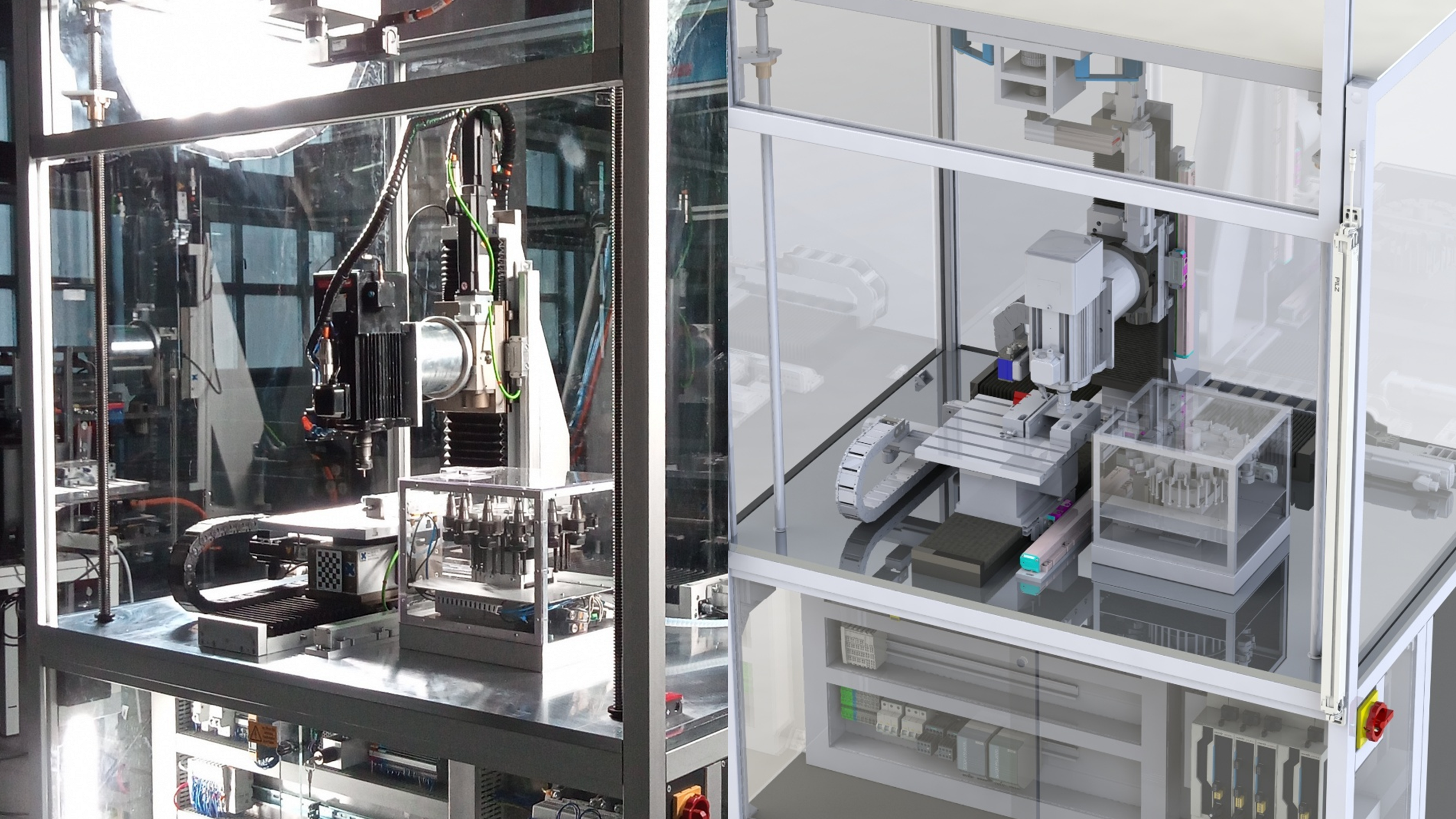}
    \caption{Five-axis milling machine (left) and its simulation model (right)}
    \label{fig:ex:fivex}
\end{figure}

To exemplify the proposed Suitability Model, we consider the engineering life cycle of a five-axis milling machine controlled by Beckhoff TwinCAT (cf. \autoref{fig:ex:fivex}).
The scenario illustrates how suitability evolve across development phases and stakeholder perspectives and how temporarily divergent digital representations can be reconciled prior to investment and operational deployment.

During the early engineering phase, the manufacturer creates a standalone AAS containing the standardized submodels ``Digital Nameplate'' and ``Technical Data''.
The technical data documents planned feed rates and spindle speeds, as well as key construction parameters derived from preliminary calculations.
Although structurally compliant with the standardized submodel templates, the values are partly estimated and have not yet been validated against finalized drive configurations.

An early customer requested the AAS machine for the planning of a future production line.
The customer assesses its suitability for capacity alignment and layout planning based on its reference AAS.
Further properties in the reference AAS are the simulation model, axes limits, and dynamics, which would enable tool path planning and validation using a material removal simulation, likewise qualified as optional.
In contrast values of the feed rates, spindle speed, and dimensions are required in the reference.
In total, this yields twenty parameters that are specified as mandatory by the submodel, nine of which are critical for capacity alignment and layout planning.
Based on \autoref{eq:suit:ref}, with nine critical values and deviations in all eleven non-critical values weighted at $w=0.2$ due to the early stage, suitability results in $S=1-(0+0.2*11)/20=0,89$. 
Based on the results, the customer expects a limited suitability, as required values are present but optional ones are not.
The available parameters allow approximate cycle time calculations, yet the absence of finalized dynamic properties and simulation models restricts advanced planning tasks such as tool path validation or material removal simulation.
During planning, the customer extends the AAS with derived production metrics, enriches the technical data, and extends its ``Process Parameters Type''.
In parallel, the manufacturer completes the detailed design of the drives, leading to slightly modified parameters in the Technical Data submodel.
Furthermore, validated engineering artifacts are integrated in ``Handover Documentation'', ``Provision of Simulation Models'', the ``Provision of 3D Models'', the ``Capability Description'' and the ``Control Component Instance'', resulting in an AAS ready to deliver with the resource.

\begin{figure}[b]
    \centering
    \includegraphics[width=0.75\linewidth]{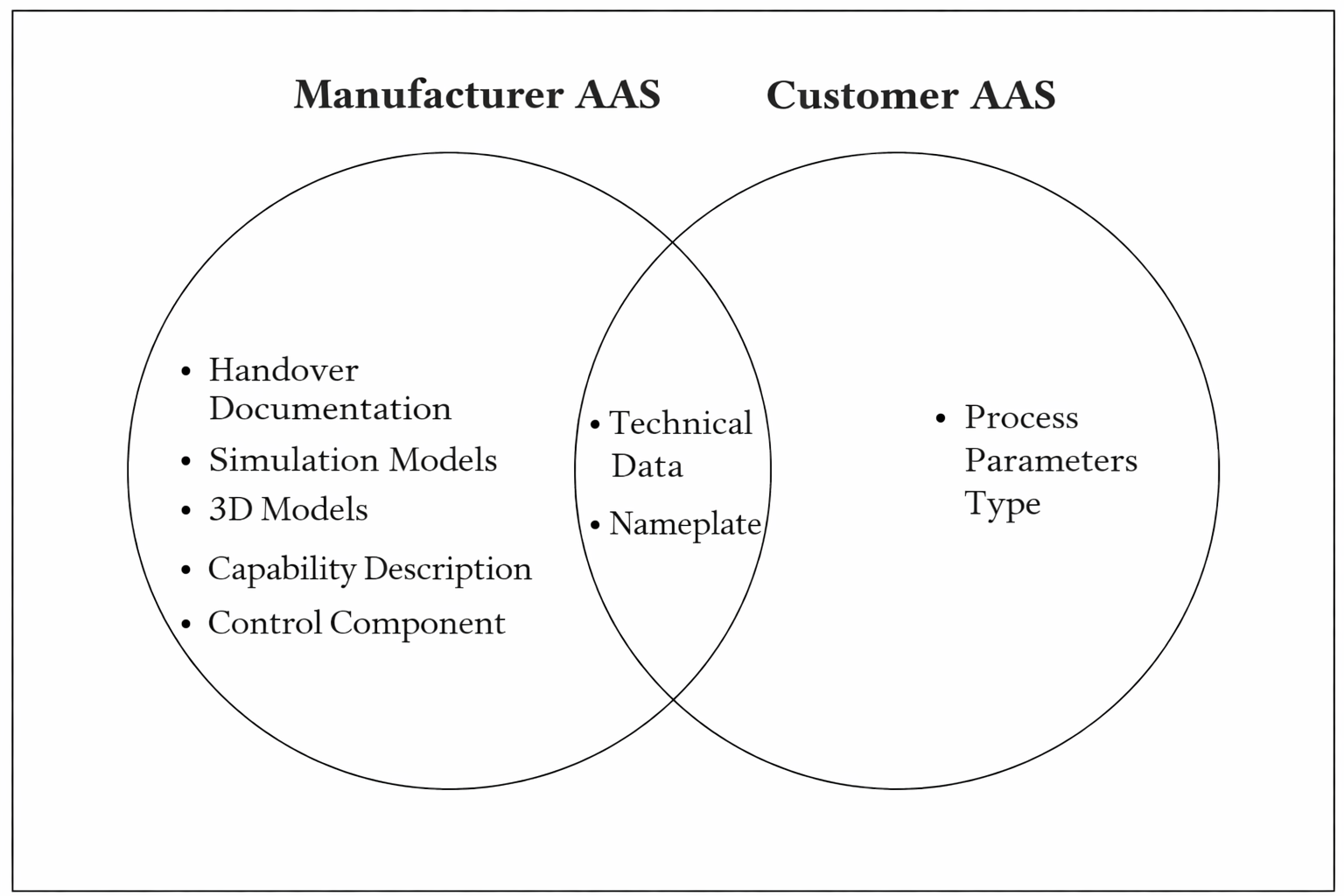} 
    \caption{Comparison of the two AAS for the milling machine using set theory.}
    \label{fig:ex:venn}
\end{figure}

At this point, two partially divergent representations exist: a customer-enhanced version and a manufacturer updated engineering version.
Before making an investment decision, the customer compares both variants.
The difference at the level of the submodels visualizing the equations defined in \autoref{sec:Comparability} is shown in \autoref{fig:ex:venn}.
The customer decides that minor deviations in the overlapping parameters remain within acceptable tolerances.
As no reevaluation of his first assessment is needed, he merges his calculations from the early development stage in the updated AAS of the manufacturer.
However, the inclusion of validated simulation models substantially increases suitability for detailed path planning, collision checks, and material removal simulation, as all deviations of non-critical parameters from the first stage are solved, the AAS is directly usable with the full suitability $S=1$.

With this exemplification, the methodologies introduced are demonstrated.
In addition to being used as a basis for suitability assessments, the comparison helps to merge divergent developments.
Based on the model, the use of the AAS in software applications is evaluated at different development stages.

\section{Limitations}
\label{sec:limitations}

The basis for compatibility in \autoref{sec:Comparability} is the simplification of the AAS structure. 
Dependencies between submodels and cross-references are not taken into account here, nor are duplicates or the order of submodels. 
In particular, with regard to the order, the applicability of set theory in large AAS containing nested elements must be verified. 
The semantic meaning of the values is ignored for the comparison based on set theory. 
For parameters with the same values but different units this could lead to misinterpretation (e.g., temperature °C or K).

The suitability in \autoref{sec:Capability} closes the gap between the pure conformity of the specification and practical usability. 
While classic validation tools check syntactic correctness, the suitability test described here addresses the application-oriented perspective.
This results in the absence of systematic data quality assessment—limited exclusively to structural aspects—and the omission of any analysis or resolution of semantic ambiguities.

\section{Conclusion}
\label{sec:Conclusion}

This paper presents methods for (i) comparing AAS models, and (ii) evaluating its suitability for a specific use case.
Their applicability is substantiated through illustrative examples.

The proposed suitability model determines the optimum for a respective use case, is thus less theoretical and more practical, and supports, in particular, the ongoing development of an AAS over time. 
If the methodology is further refined by means of the following steps and its limitations are systematically mitigated, it may provide an additional basis for the comparative analysis of AAS and thereby augment the set-theoretic comparison methodology previously presented.  
Such an extension would enable the parallelized and collaborative development of assets within the context of AAS.  
In particular, it would support the realization of a Git-like branching and merging mechanism, which would constitute a highly valuable instrument for engineering activities as well as for life cycle management.

\section*{Acknowledgments}
Partly funded by the German Federal Ministry of Research, Technology and Space (BMFTR) within the ``Research Campus – Public-Private Partnership for Innovation'' funding initiative (grant no. 02P23Q820) and managed by the Project Management Agency Karlsruhe (PTKA).
Partly funded by the Federal Ministry for Economic Affairs and Energy (BMWE) through the projects growING (grant no. 13IPC036G).
\bibliography{temp}
\bibliographystyle{elsarticle-num}


\normalMode
\end{document}